\documentclass[
  aps,
  pra,
  twocolumn,
  showpacs,
  superscriptaddress,
  floatfix
]{revtex4-2}
\usepackage[dvipdfmx]{graphicx}
\usepackage{verbatim}
\usepackage{epsf}
\usepackage{natbib}
\usepackage{dcolumn}
\usepackage{bm}
\usepackage{color} 
\usepackage{ulem}
\usepackage{amsmath}
\usepackage{amssymb}
\usepackage{txfonts}
\usepackage{tikz}
\usetikzlibrary{matrix,arrows,decorations.pathmorphing}
\usepackage[
  colorlinks=true,
  citecolor=blue,
  urlcolor=blue,
  setpagesize=false]{hyperref}
\usepackage{nicematrix}

\newcommand{\Tr}  {{\rm Tr}}

\newcommand{\pc}{p_{\mathrm{c}}}
\newcommand{\Lmin}{L_{\mathrm{min}}}
\newcommand{\Lmax}{L_{\mathrm{max}}}

\hypersetup{
  colorlinks=true,
  linkcolor=[rgb]{0.6,0.0,0.0},
  citecolor=[rgb]{0.0,0.0,0.6},
  urlcolor=[rgb]{0.0,0.0,0.6},
  setpagesize=false
}

\begin{document}

\title{
  Quantum conditional mutual information as a probe of measurement-induced entanglement phase transitions
}

\author{Yuichi~Otsuka}
\email{otsukay@riken.jp}
\affiliation{Quantum Computational Science Research Team, RIKEN Center for Quantum Computing (RQC), Saitama 351-0198, Japan}
\affiliation{Computational Materials Science Research Team, RIKEN Center for Computational Science (R-CCS),  Hyogo 650-0047,  Japan}

\author{Kazuhiro~Seki}
\email{kazuhiro.seki@riken.jp}
\affiliation{Quantum Computational Science Research Team, RIKEN Center for Quantum Computing (RQC), Saitama 351-0198, Japan}

\author{Seiji~Yunoki}
\email{yunoki@riken.jp}
\affiliation{Quantum Computational Science Research Team, RIKEN Center for Quantum Computing (RQC), Saitama 351-0198, Japan}
\affiliation{Computational Materials Science Research Team, RIKEN Center for Computational Science (R-CCS),  Hyogo 650-0047,  Japan}
\affiliation{Computational Quantum Matter Research Team, RIKEN Center for Emergent Matter Science (CEMS), Saitama 351-0198, Japan}
\affiliation{Computational Condensed Matter Physics Laboratory, RIKEN Cluster for Pioneering Research (CPR), Saitama 351-0198, Japan}

\begin{abstract}
  
  We propose that the quantum conditional mutual information (QCMI), 
  computed with a suitably chosen partition of the system, serves as 
  a powerful probe
  for detecting measurement-induced entanglement phase transitions in monitored quantum circuits. 
  To demonstrate this, we investigate monitored variable-range 
  Clifford circuits and identify the phase boundary
  between volume-law and area-law entanglement phases 
  by performing finite-size scaling analyses of the QCMI. 
  Assuming that the entanglement entropy exhibits a logarithmic 
  dependence on system size at criticality in short-range 
  interacting cases, we further show that the QCMI allows for 
  the simultaneous determination of both the critical point and 
  the universal coefficient of the logarithmic term in the 
  entanglement entropy via a crossing-point analysis. 
  For the shortest-range interacting case studied, we obtain the 
  thermodynamic-limit value of the coefficient as 
  $\tilde{c}=1.519(3)$, which is significantly 
  smaller than values reported in previous studies. 
\end{abstract}

\date{\today}

\maketitle

\section{Introduction}\label{sec:intro}

Measurement-induced phase transitions (MIPTs)
in quantum circuits composed of unitary dynamics and
local projective measurements have attracted significant interest 
in recent years~\cite{Li-PRB2018,Li-PRB2019a,Chan-PRB2019,Skinner-PRX2019,Fisher-ARCMP2023}.
These transitions separate two distinct entanglement phases: 
a volume-law phase, in which the entanglement entropy of a subsystem 
scales proportionally with its volume, and an 
area-law phase, where it scales 
with the boundary area of the subsystem.
Such transitions emerge from the competition between unitary 
dynamics, which generally generate entanglement among qubits, and
local projective measurements, which tend to disentangle the measured 
qubits from the rest of the system. 
Remarkably, measurement-induced entanglement phase 
transitions~\cite{Noel-NatPhys2022,Koh-NP2023}, as well as other nonequilibrium 
phase transitions~\cite{Chertkov_NatPhys2023}, have been demonstrated 
experimentally on real quantum devices. 
These experimental breakthroughs have been made possible either by 
employing mid-circuit measurement and reset 
functionalities~\cite{Koh-NP2023,Chertkov_NatPhys2023}, or by 
using a reference qubit initially maximally entangled with the monitored 
system as a probe for the entanglement phase 
transition~\cite{Noel-NatPhys2022,Gullans-PRX2020}. 

To numerically detect the measurement-induced entanglement phase 
transitions,
the entanglement entropy itself is not always a suitable quantity, 
particularly due to its emergent logarithmic dependence on subsystem size 
at criticality in (1+1)$d$ systems~\cite{Skinner-PRX2019,Li-PRB2019a}.
Indeed, performing finite-size scaling with entanglement entropy requires 
subtracting its value at an estimated critical point 
(e.g., a critical measurement rate) to eliminate the logarithmic 
dependence~\cite{Skinner-PRX2019}, which introduces an additional source 
of uncertainty. 
As an alternative, 
the tripartite mutual information (TMI)~\cite{Kitaev-PRL2006}, 
defined as  
$I_{3}(A,B,C)=S_{A}+S_{B}+S_{C}-S_{A\cup B}-S_{B\cup C}-S_{A\cup C}-S_{A\cup B \cup C}$,  
where $S_X$ is the entanglement entropy of subsystem $X$, 
has been used to prove MIPTs~\cite{Gullans-PRX2020,Zabalo-PRB2020}. 
With an appropriately chosen tripartiton $A,B,$ and $C$, this linear 
combination of entanglement entropies effectively distinguishes 
volume-law and area-law phases 
without being affected by the logarithmic dependence at criticality.  
This approach has proven successful in systems with 
variable-range (power-law decaying) interactions~\cite{Minato-PRL2022,Block-PRL2022,Muller-PRL2022,Sahu-PRB2022} as well as in systems 
with sparse nonlocal interactions~\cite{Hashizume-PRR2022}.
While the effectiveness of the TMI is empirically established, 
it remains unclear--at least to our knowledge--why the logarithmic 
contribution is suppressed, 
especially given that $I_{3}(A,B,C)$ involves an odd number (seven) 
of entanglement entropy terms.

In this paper, we propose that 
another linear combination of entanglement entropies, of the form~\cite{Levin-PRL2006},
\begin{equation}
  I(A:C|B) =S_{A\cup B}+S_{B \cup C}-S_{B}-S_{A \cup B \cup C}, 
  \label{eq:cond}
\end{equation}
with an appropriately chosen partition $A,B,$ and $C$ of the system,
serves as an effective and pertinent probe of MIPTs.
The quantity $I(A:C|B)$ is known as the quantum conditional mutual 
information (QCMI)~\cite{Cerf-PRL1997,Cerf-PhysicaD1998}, 
and has been recently employed to quantify 
measurement-induced 
topological phase transitions~\cite{Lavasani-NatPhys2021,Lavasani-PRL2021,Kells-SciRep2023},
and
entanglement growth in stabilizer circuits composed of nearest-neighbor 
two-qubit Clifford gates~\cite{Sang-PRXQ2023}. 
The QCMI has also been used to characterize one-dimensional topological insulators and superconductors at equilibrium~\cite{Maiellaro-PRR2022,Maiellaro-PRB2023}.

Here, we demonstrate that the QCMI is also well-suited for probing 
measurement-induced entanglement phase transitions, through numerical 
simulations of one-dimensional variable-range ($1/r^\alpha$) Clifford 
circuits interspersed with local projective measurements. 
While similar models have been studied 
previously~\cite{Block-PRL2022,Sahu-PRB2022}, 
our circuit definition differs slightly in detail. 
We show that the QCMI is free from multiplicative and additive renormalization; at criticality, it becomes system-size independent under suitable partitioning of the system. This property makes the QCMI particularly useful for identifying critical points and extracting universal quantities much like the Binder ratio in conventional statistical physics. Notably, the value of the QCMI at emergent conformal criticality directly yields the universal coefficient of the logarithmic term.
Through careful finite-size-scaling analyses of $I(A:C|B)$, 
we find that no phase transition occurs for $\alpha \lesssim 2$, 
whereas for $2 \lesssim \alpha \lesssim 3$, measurement-induced 
entanglement phase transitions emerge with critical exponents 
distinct from those in nearest-neighbor circuits and in the regime 
$3 \lesssim \alpha$. 
These findings are qualitatively consistent with earlier studies
on monitored circuits with variable-range interactions, not restricted to 
Clifford circuits~\cite{Minato-PRL2022,Block-PRL2022,Muller-PRL2022,Sahu-PRB2022}.
Furthermore, using crossing-point analysis of $I(A,C|B)$, we extract 
the coefficient $\tilde{c}$ of the emergent logarithmic scaling of the 
entanglement entropy at criticality in the thermodynamic limit. 
For the shortest-range interacting case studied, we find 
$\tilde{c}=1.519(3)$, which is significantly smaller than 
previous estimates obtained in monitored nearest-neighbor random 
Clifford circuits~\cite{Li-PRB2019a,Zabalo-PRB2020,Sierant-PRB2022}.

The rest of this paper is organized as follows. 
In Sec.~\ref{sec:model}, we define the stabilizer circuit with 
variable-range two-qubit Clifford gates and 
illustrate how the QCMI behaves at criticality under a suitable 
partition of the system.
In Sec.~\ref{sec:results}, we numerically demonstrate that
the QCMI not only detects the measurement-induced entanglement 
transition, but also enables estimation of the coefficient of 
the logarithmic term in the entanglement entropy at criticality.  
In Sec.~\ref{sec:discussion}, we discuss the correlation-length critical 
exponent and the universal coefficient of the logarithmic term, and 
compare them with values obtained in previous studies. 
Finally, we present conclusions and remarks in 
Sec.~\ref{sec:conclusion}.

\section{Model and method}\label{sec:model}

\subsection{Stabilizer circuit}

We investigate entanglement phase transitions in stabilizer circuits 
composed of variable-range two-qubit Clifford gates and local projective 
measurements. 
Stabilizer circuits are well suited for our purposes, 
as they enable efficient simulation of the dynamics of thousands of 
qubits, not only in weakly entangled states 
but also in highly entangled ones. This stands 
in contrast to exact diagonalization or tensor network state 
approaches, which are typically limited to systems with smaller system 
sizes or low entanglement.

\begin{figure}
  \includegraphics[width=1.0\columnwidth]{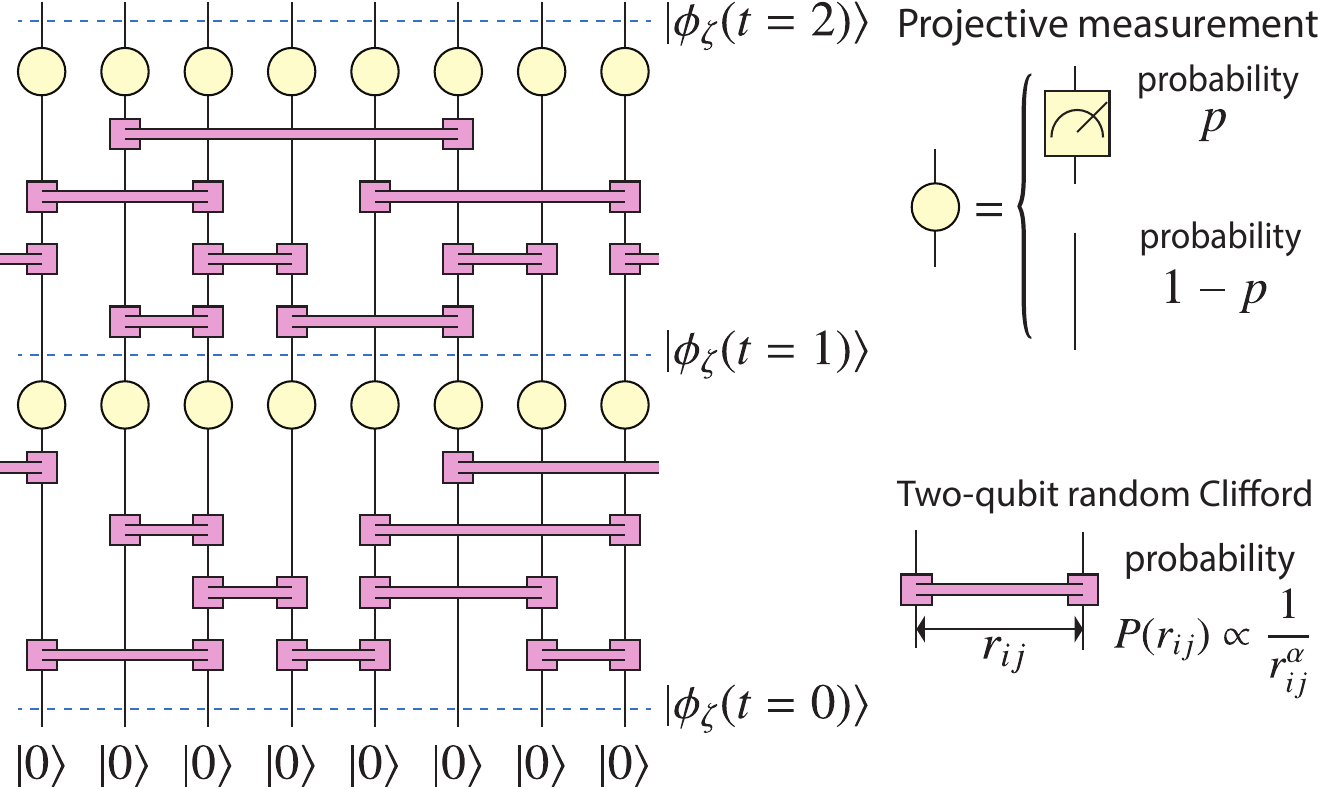}
  \caption{
    \label{fig:schematic}
    Schematic figure of a monitored variable-range random Clifford 
    circuit. 
    Time flows from bottom to top. 
    Vertical lines represent qubits, 
    dumbbell-shaped symbols indicate two-qubit random Clifford 
    gates, and 
    circles denote projective measurements that occur probabilistically. 
  }
\end{figure}

We consider a one-dimensional chain of $L$ qubits with periodic 
boundary conditions, assuming that $L$ is even. 
Let $|\phi_{\zeta}(t)\rangle$ denote the pure quantum state of the 
entire system of $L$ qubits at discrete time step 
(simply referred to as ``time'') $t$,  
initialized as $|\phi_{\zeta}(t=0)\rangle=|0\rangle^{\otimes L}$. 
Here, $\zeta$ indexes different circuit realizations characterized by 
random seeds. 
Time $t$ is incremented by one after each application of $L$ two-qubit 
Clifford gates, followed by projective measurements on each qubit 
along the Pauli-$Z$ axis with probability $p$ 
(see dashed horizontal lines in Fig.~\ref{fig:schematic}).  
Each two-qubit Clifford gate is randomly selected from the complete set 
of 11520 distinct two-qubit Clifford 
gates~\cite{Corcoles-PRA2013,Richter-PRR2023}, 
and applied to a randomly chosen pair of qubits $i$ and $j(\not{=}i)$ 
with probability
\begin{equation}
  P(r_{ij})=\frac{1}{\cal N} \frac{1}{r_{ij}^\alpha}, 
\end{equation}
where $r_{ij}$ is the distance between qubits $i$ and $j$
under periodic boundary conditions, defined as   
$r_{ij}=|i-j|$ for $|i-j|\leqslant L/2$ and
$r_{ij}=L-|i-j|$ for $|i-j|> L/2$, 
${\cal N}=\sum_{x=1}^{L/2}\frac{1}{x^\alpha}$ is the normalization 
constant, and $\alpha$ is a parameter that controls the interaction 
range of the two-qubit gates.
We simulate circuit dynamics using the tableau 
algorithm~\cite{Aaronson-PRA2004} 
(see also Supplemental Material~\cite{suppl}), for system sizes 
ranging from $L=16$ to $2048$, and evolution times up to 
$t_{\rm max}=4096$.
The parameters $\alpha$ and $p$ are varied in the ranges 
$1 \leqslant \alpha \leqslant 4$ and $0 \leqslant p < 1$, respectively.

\subsection{Entanglement entropy and quantum conditional mutual information}

\begin{figure}
  \includegraphics[width=1.0\columnwidth]{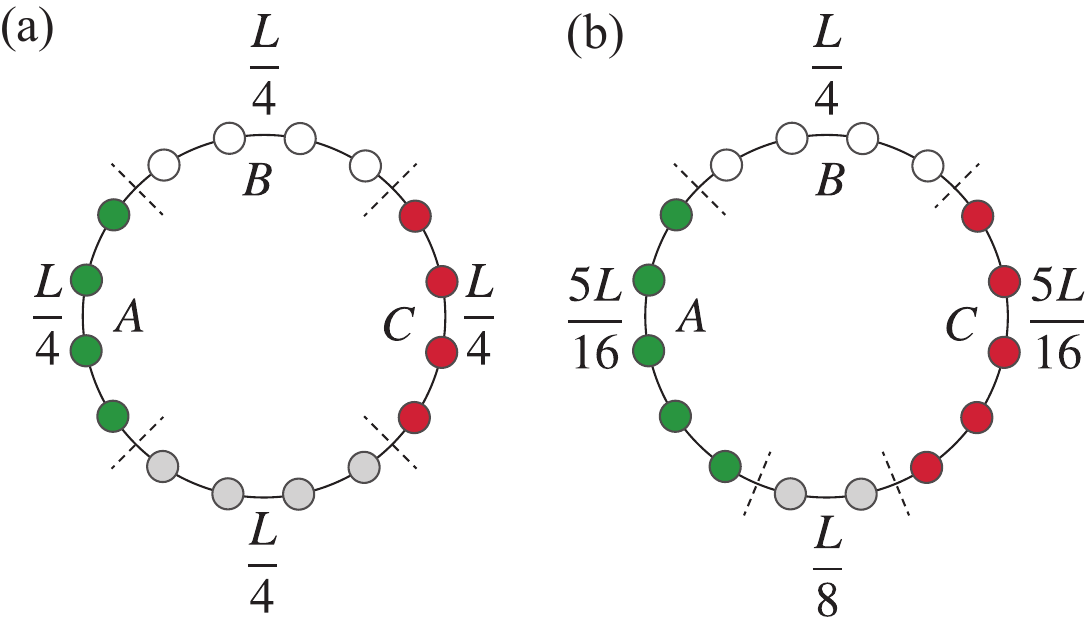}
  \caption{
    \label{fig:ABC}
    Two distinct partitions of the $L$-qubit system into
    subsystems $A$ (green), $B$ (white), $C$ (red), and 
    $\overline{A\cup B\cup C}$ (gray).  
    Each circle represents a qubit, and the number of qubits 
    in each segment is indicated. Partition (a) is used for 
    the QCMI in Eq.~(\ref{eq:IACBa}), while partition (b) is employed for the QCMI in Eq.~(\ref{eq:IACBb}).   
  }
\end{figure}

At each time step $t$, we compute the von Neumann entanglement 
entropies of the state $|\phi_{\zeta}(t)\rangle$ for subsystems $X= A\cup B, B \cup C, B$, and $ A \cup B \cup C$
(see Fig.~\ref{fig:ABC}) using the base-2 logarithm:  
\begin{equation}
  S_{X,\zeta}(t) =-\Tr_{X}\left[\hat{\rho}_{X,\zeta}(t) \log_2 \hat{\rho}_{X,\zeta}(t)\right],
\end{equation}
where $\hat{\rho}_{X}(t)=\Tr_{\overline{X}}\left[ |\phi_{\zeta}(t)\rangle \langle \phi_{\zeta}(t)| \right]$
is the reduced density matrix of subsystem $X$, 
and $\bar X$ denotes its complement. 
To compute the entanglement entropy in stabilizer circuits,
we follow the method described in Ref.~\cite{Nahum-PRX2017} 
(see also Supplemental Material~\cite{suppl}). 
%
Note that when $p=1$, the entanglement entropy vanishes for all 
subsystems since the state $|\phi_{\zeta}(t)\rangle$ becomes a 
product state  immediately after every measurement. 

After sufficiently long time evolution, the state 
$|\phi_\zeta(t)\rangle$ reaches a steady state in which 
the entanglement entropy $S_{X,\zeta}(t)$ fluctuates around 
a saturation value.
Such a steady state can exhibit a phase transition between 
volume-law and area-law phases as 
the parameter $p$ or $\alpha$ is varied, 
in the thermodynamic limit ($L\to \infty$). 
From the entanglement entropy $S_{X,\zeta}(t)$, 
we evaluate the steady-state QCMI as 
\begin{equation}
  I(A:C|B) = \frac{1}{N_t}\sum_{k=0}^{N_{t}-1}
  \left \langle \left \langle I_{\zeta}(t_{\rm min} + k \Delta_{I,t}) \right \rangle \right\rangle, 
\end{equation}
where 
$I_{\zeta}(t) \equiv S_{A\cup B,\zeta}(t)+S_{B \cup C,\zeta}(t)-S_{B,\zeta}(t)-S_{A \cup B \cup C,\zeta}(t)$
is the QCMI for the state $|\phi_{\zeta}(t)\rangle$. 
Here, $t_{\rm min}$ is the time after which 
$I_{\zeta}(t)$ has reached saturation on average,  
$\Delta_{I,t}$ is the sampling time interval,
$N_t$ is the number of time steps used in the time average, 
and $\langle \langle \cdots \rangle \rangle \equiv \frac{1}{R}\sum_{\zeta=1}^R \cdots$
denotes the statistical average over $R$ independent circuit 
realizations.
We set
$t_{\rm min}=\min(2L,2048)$ and $R=10560$, and determine 
$\Delta_{I,t}$ as the estimated autocorrelation time of
$\left \langle \left \langle I_{\zeta}(t) \right \rangle \right\rangle$ for $t \geqslant t_{\rm min}$, 
assuming exponential decay of the corresponding autocorrelation 
function. 
In the worst case, near criticality for $L=2048$, 
$\Delta_{I,t} \sim 100$, while it is significantly 
smaller away from criticality. 
$N_t$ is chosen as the maximum integer satisfying 
$t_{\rm min} + (N_t-1) \Delta_{I,t} \leqslant t_{\rm max}$.

\subsection{Entanglement entropy at criticality}

Motivated by the emergent conformal criticality  
observed in monitored Clifford and non-Clifford circuits with 
short-range interactions or hoppings~\cite{Skinner-PRX2019,Li-PRB2019a,Fuji-PRB2020,Alberton-PRL2021,Minato-PRL2022,Block-PRL2022}, 
we assume that the steady-state entanglement entropy 
$S_{X}=\frac{1}{N_t}\sum_{k=0}^{N_t-1}
\langle \langle S_{X,\zeta}(t_{\rm min} + k \Delta_{S_X,t}) \rangle \rangle $
at criticality behaves as~\cite{Calabrese-JPAMT2009}
\begin{equation}
  S_{X}\sim\frac{c}{3} \log_2{l_X} +c' \quad (p = p_{\rm c}),
  \label{eq:Sx}
\end{equation}
where $l_X\equiv \frac{L}{\pi}\sin\frac{\pi|X|}{L}$ is the chord 
length of subsystem $X$ with $|X|$ denoting the number of qubits 
in $X$, 
$p_{\rm c}$ the critical value of the measurement rate, 
$c$ a universal coefficient of the logarithmic term 
(often referred to as an effective central 
charge)~\cite{Jian-PRB2020}, and $c'$ a nonuniversal constant.
The autocorrelation time $\Delta_{S_X,t}$ for the steady-state 
entanglement entropy is estimated in the same manner as 
$\Delta_{I,t}$, and exhibits similar behavior.
Notice that the coefficient $c$ here is not equivalent to the 
central charge in conformal field theory. 
For example, in the case of the percolation universality class, 
the central charge is known to be 
zero~\cite{Jian-PRB2020,Oshima-PRB2023}.
Note also that in the deep volume-law phase 
($p \ll p_c$), the entanglement entropy scales 
extensively as $S_{X}\sim |X|$, while in the area-law phase 
($p \gg p_{\rm c}$), it scales with the boundary, 
$S_X \sim |\partial X|$, where 
$|\partial X|$ denotes the number of qubits on the boundary of 
subsystem $X$.

\subsection{Quantum conditional mutual information at criticality}

To use the QCMI as a probe for the measurement-induced entanglement 
phase transition, 
we partition the system according to the following three conditions 
(see also Fig.~\ref{fig:ABC}): 
(i) The system is divided into four parts 
$A, B, C,$ and $\overline{A\cup B \cup C}$, where 
$A, B,$ and $C$ are contiguous segments located in the middle of the
periodic chain. 
This setup ensures that each entanglement entropy $S_X$ is 
associated with a different chord length $l_{X}$, while all 
share the same nonuniversal constant $c'$. 
By substituting Eq.~(\ref{eq:Sx}) into the definition of QCMI in 
Eq.~(\ref{eq:cond}), we obtain 
\begin{equation}
  I(A:C|B) \sim 
  \frac{c}{3} \log_2
  \frac
      {l_{A\cup B} l_{B\cup C}}
      {l_{B} l_{A\cup B\cup C}}
      \quad (p = p_{\rm c}),
  \label{eq:IACB0}
\end{equation}
indicating that the nonuniversal constant $c'$ cancels out in 
$I(A:C|B)$ at criticality. 
(ii) To eliminate the dependence of $I(A:C|B)$ on the total system 
size $L$ at criticality, 
we choose the subsystem sizes such that $|A\cup B|$, $|B \cup C|$, 
$|B|$, and $|A \cup B \cup C|$ are all proportional to $L$.
(iii) To capture the volume-law entanglement phase, the subsystem 
sizes are chosen such that 
$L_{A\cup B}+L_{B \cup C} \ne L_{B}+L_{A \cup B \cup C}$, 
because for $p \ll p_{\rm c}$, the QCMI behaves as 
$I(A:C|B)\sim L_{A\cup B}+L_{B \cup C} - L_{B} - L_{A \cup B \cup C}$ , 
where $L_X \equiv \min(|X|,L-|X|)$ is the smaller of the number of 
qubits in $X$ and its complement $\overline{X}$. 
We note that in the area-law phase $p \gg p_{\rm c}$, the QCMI 
behaves as $I(A:C|B)\sim 0$, regardless of the partitioning. 
Moreover, the inequality $I(A:C|B)\geqslant 0$ always holds due to 
the strong subadditivity of von Neumann entanglement 
entropy~\cite{Lieb-PRL1973}.
Thus, as long as the partitioning satisfies conditions (i)-(iii), 
the QCMI $I(A:C|B)$ serves as a meaningful probe of the 
measurement-induced entanglement phase transition.
Importantly, condition (iii) is independent of the assumption of 
conformal criticality in Eq.~(\ref{eq:Sx}), 
implying that $I(A:C|B)$ can detect the entanglement phase 
transition even in the absence of emergent conformal invariance.

Specifically, we consider two types of partitioning for 
the same system, in a spirit of Ref.~\cite{Furukawa-PRL2009}. 
The first partitioning sets 
$|A|=|B|=|C|=L/4$ [see Fig.~\ref{fig:ABC}(a)], for which we find  
\begin{equation}
  I_a(A:C|B) \sim \frac{c}{3} 
  \log_2{\frac{\sin^2{\frac{\pi}{2}}}{\sin{\frac{\pi}{4}}\sin{\frac{3\pi}{4}}}}
    =\frac{c}{3}  \quad \quad (p = p_{\rm c}).
  \label{eq:IACBa}
\end{equation}
The second partitioning is given by 
$|A|=|C|=\frac{5L}{16}$ and $|B|=\frac{L}{4}$ 
[see Fig.~\ref{fig:ABC}(b)], for which we obtain 
\begin{equation}
  I_b(A:C|B) \sim \frac{c}{3} \log_2{\frac{\sin^2{\frac{9\pi}{16}}}{\sin{\frac{\pi}{4}} \sin{\frac{\pi}{8}}}}
  \approx 1.83 \times \frac{c}{3}
  \quad \quad (p = p_{\rm c}).
  \label{eq:IACBb}
\end{equation}
Here, the subscripts $a$ and $b$ in $I_a$ and $I_b$ correspond to 
the partitions illustrated in Figs.~\ref{fig:ABC}(a) and 
\ref{fig:ABC}(b), respectively. 
It should be emphasized that Eqs.~(\ref{eq:IACB0})--(\ref{eq:IACBb}) 
are valid only under the assumption of logarithmic scaling of the 
entanglement entropy at criticality, as expressed in 
Eq.~(\ref{eq:Sx}). 
As we show in the next section, our numerical results strongly 
support the validity of this assumption for $\alpha \gtrsim 3$.

\section{Results}\label{sec:results}

\subsection{Phase diagram}
\label{subsec:phase_diagram}

\begin{figure}
  \includegraphics[width=1.0\columnwidth]{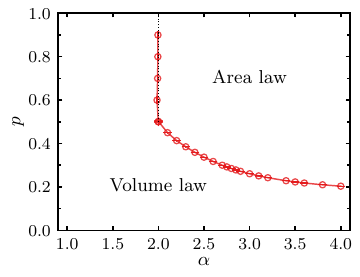}
  \caption{
    \label{fig:phasediagram}
    Phase diagram in the thermodynamic limit. 
    The phase boundary is estimated from a data-collapse analysis 
    of the QCMI $I_a(p, L)$, evaluated using the partitioning 
    scheme shown in Fig.~\ref{fig:ABC}(a). An equivalent phase 
    diagram is obtained using the QCMI $I_b(p, L)$ based on 
    the alternative partitioning shown in Fig.~\ref{fig:ABC}(b).  
    The dashed vertical line at $\alpha=2$ is a guide to the eye. 
  }
\end{figure}

Before presenting a detailed analysis, we first show the global 
phase diagram of the stabilizer circuits with variable-range 
two-qubit Clifford gates as a function of $\alpha$ and 
$p$ in Fig.~\ref{fig:phasediagram}.
Using the QCMI as a probe, we accurately determine the phase 
boundary separating the area-law and volume-law entanglement phases 
for $\alpha > 2$.
In what follows, we analyze the critical properties of the system 
along this phase boundary. 
Additionally, we confirm that no area-law phase exists for 
$\alpha \lesssim 2$,
which is consistent with the previous analytical arguments in 
the high-measurement-rate regime~\cite{Block-PRL2022,Sahu-PRB2022}.

\subsection{Quantum conditional mutual information near criticality}
\label{subsec:poorman}

\begin{figure}
 \includegraphics[width=1.0\columnwidth]{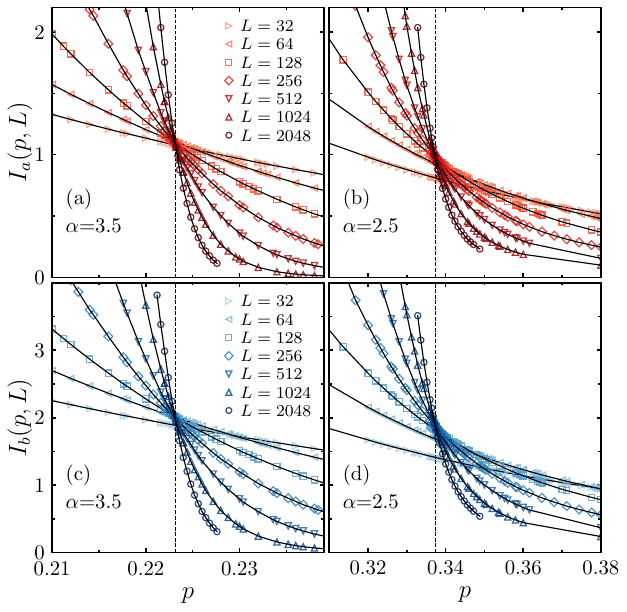}
 \caption{
 \label{fig:I-p}
 QCMI $I_\gamma(p, L)$ for different partitions and interaction 
 ranges. Panels (a) and (b) show results for $\gamma=a$, while  
 (c) and (d) correspond to $\gamma=b$. 
 The interaction-range parameter is $\alpha=3.5$ for (a) and (c), 
 and $\alpha=2.5$ for (b) and (d). 
 Vertical dashed lines indicate the critical points estimated 
 from the data-collapse analyses in Figs~\ref{fig:vsLmin-alpha3500} 
 and \ref{fig:vsLmin-alpha2500}.
 }
\end{figure}

Let us denote the QCMI as $I_{\gamma}(p,L)$, where $\gamma=a$ 
and $b$ correspond to the two distinct partitioning schemes shown 
in Figs.~\ref{fig:ABC}(a) and \ref{fig:ABC}(b), respectively. 
At the critical point $p=\pc$, the QCMI becomes independent of the 
system size [see Eqs.~(\ref{eq:IACBa}) and~(\ref{eq:IACBb})], 
which suggests that the crossing-point analysis~\cite{Luck-PRB1985} 
can be effectively used to estimate both $\pc$ and the 
correlation-length critical exponent $\nu$. 
Indeed, as shown in Fig.~\ref{fig:I-p}, the curves of 
$I_{\gamma}(p, L)$ for different $L$ appear to cross 
at a single point for sufficiently large system sizes. 
The estimated critical points $\pc$ are nearly identical for 
both partitions $\gamma=a$ and $b$, 
for each value of $\alpha$. This already demonstrates, at least 
in part, that the QCMI serves as a reliable probe to detect the 
critical point. 
For smaller values of $\alpha$, deviations from the size-independent 
behavior are more pronounced in smaller systems  
[see Figs.~\ref{fig:I-p}(b) and \ref{fig:I-p}(d)]. 
Nevertheless, well-established methods for handling corrections to 
scaling have been developed and documented in 
the literature~\cite{Shao-S2016}.
Interestingly, we find that corrections to scaling becomes more 
subtle for larger $\alpha$. 
As discussed in Sec.~\ref{subsec:crossing}, 
nonmonotonic finite-size behavior of the crossing points 
[see Figs.~\ref{fig:px-L}(a) and \ref{fig:Ix-L}(a)] 
emerges in the vicinity of the critical point, 
suggesting that multiple correction terms may need to be 
considered. A similar issue was only recently resolved 
in the study of the N\'{e}el-paramagnetic quantum phase transition 
in dimerized Heisenberg models, where the known value of the 
correlation-length exponent $\nu$ from the expected O(3) 
universality class was crucial for the analysis~\cite{Ma-PRL2018}.
In our case, however, neither the critical exponent $\nu$ nor 
the universality class is known a priori, necessitating the use of 
an alternative approach, as described in the following.

\subsection{Data-collapse}
\label{subsec:collapse}

\begin{figure}
 \includegraphics[width=0.975\columnwidth]{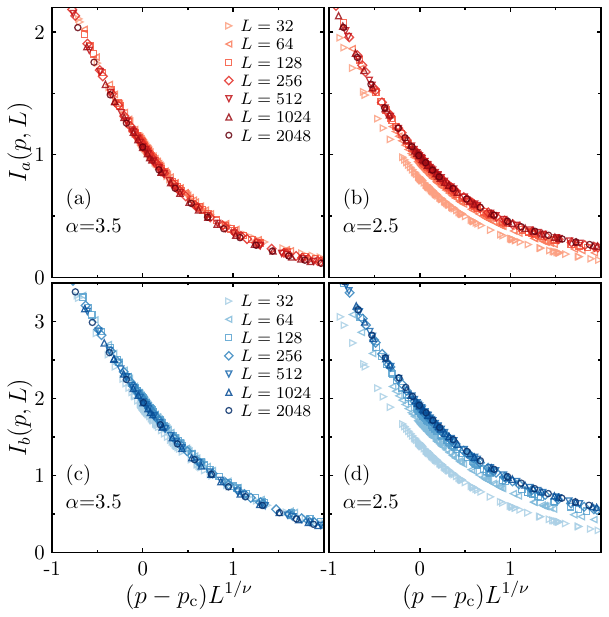}
 \caption{
 \label{fig:I-x}
 Data-collapse fits of the QCMI $I_\gamma(p, L)$.  
 Panels (a) and (b) show results for $\gamma=a$, while 
 (c) and (d) correspond to $\gamma=b$.  
 The interaction-range parameter is $\alpha=3.5$ for (a) and (c), and $\alpha=2.5$ for (b) and (d).
 The extrapolated values of $\pc$ and $\nu$ in the thermodynamic 
 limit, as obtained in Figs~\ref{fig:vsLmin-alpha3500} 
 and \ref{fig:vsLmin-alpha2500}, are used here.
 }
\end{figure}

\begin{figure}
 \centering
 \includegraphics[width=0.95\columnwidth]{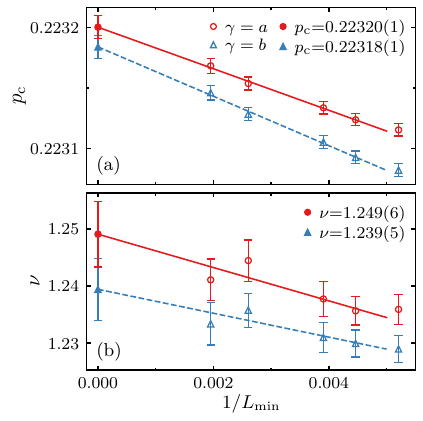}
 \caption{
 \label{fig:vsLmin-alpha3500}
 Extrapolation of (a) $\pc$ and (b) $\nu$ obtained from 
 data-collapse analysis using the QCMI $I_{\gamma}(p,L)$ for 
 $L \geqslant \Lmin$ at $\alpha=3.5$. 
 Solid and dashed lines represent linear fits to the data for 
 $\gamma=a$ and $\gamma=b$, respectively. 
 The extrapolated values in the thermodynamic limit 
 ($1/\Lmin\to0$) are indicated by solid symbols at $1/\Lmin=0$.
 }
\end{figure}

\begin{figure}
 \centering
 \includegraphics[width=0.95\columnwidth]{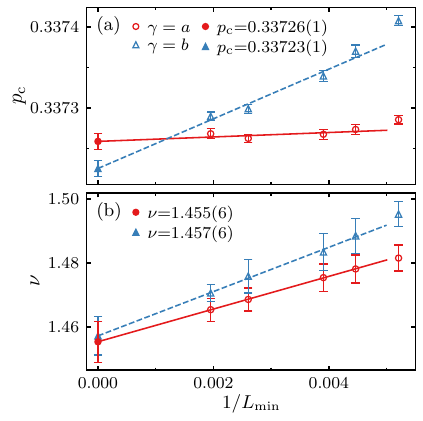}
 \caption{
 \label{fig:vsLmin-alpha2500}
 Same as Fig.~\ref{fig:vsLmin-alpha3500}, but for $\alpha=2.5$.
 }
\end{figure}

Here, we adopt an empirical approach to 
estimate $\pc$ and $\nu$ in the thermodynamic limit 
based on a data-collapse analysis.
We employ the Bayesian scaling analysis method~\cite{Harada-PRE2011} 
to collapse the data of $I_{\gamma}(p, L)$, 
assuming the following scaling ansatz: 
\begin{equation}
 I_\gamma(p,L) = f_\gamma( \Delta p L^{1/\nu} ),
\end{equation}
where $\Delta p = p - \pc$ is the deviation from the critical point, 
and $f_{\gamma}$ denotes the scaling function.
Representative examples of successful data-collapse fits are shown 
in Fig.~\ref{fig:I-x}.
We do not explicitly include a correction term of the form
$\lambda L^{-\omega}$ in the scaling function, 
where $\lambda$ is an irrelevant scaling field and $\omega$ is 
the associated correction exponent.
Instead, we perform data-collapse fits over the system size range 
$L \in [\Lmin, \Lmax]$ with $\Lmax=2048$ fixed and $\Lmin$ 
systematically varied. We then examine 
the asymptotic behavior of the estimated values of $\pc$ and $\nu$ 
as $\Lmin$ increases~\cite{Otsuka-PRX2016,Otsuka-PRB2020}.
An important aspect of this analysis is the estimation of 
uncertainties in the fitted parameters. 
Because data collapsing is inherently a nonlinear procedure, 
the results are sensitive to the choice of initial parameters 
in the Bayesian analysis as well as the selected data set.
To mitigate these sensitivities and obtain reliable error bars, we 
apply the bootstrap resampling method. 
The resulting uncertainties of the estimated $\pc$ and $\nu$ are 
shown 
in Figs.~\ref{fig:vsLmin-alpha3500} and \ref{fig:vsLmin-alpha2500}. 
We extrapolate the results to the thermodynamic limit, i.e., 
$1/\Lmin=0$, using a linear fit based on the four largest values of 
$\Lmin$. Although the use of a linear fit and the specific number 
of data points are empirical choices, we find 
that the estimated values of $\pc$ and $\nu$ in the thermodynamic 
limit are consistent between $\gamma=a$ and $\gamma=b$ within two 
standard deviations. 
As summarized in Table~\ref{tbl:pc-nu},
the results obtained for other values of $\alpha$ 
are overall consistent between the two different partitioning 
schemes, agreeing to at least four (three) significant digits for 
$\pc$ ($\nu$) in most cases of $\alpha$. 
Additional analyses are provided in Supplemental Material~\cite{suppl}.

\begin{table}[tb]
 \caption{\label{tbl:pc-nu}%
 Summary of the critical point $\pc$ and the correlation-lemgth 
 exponent $\nu$ in the thermodynamic limit, 
 estimated from data-collapse analysis of the QCMI $I_{a}$ 
 and $I_{b}$ for the two different partitioning schemes. 
 The numbers in parentheses indicate the uncertainty 
 in the last digits.
 }
 \begin{ruledtabular}
  \begin{tabular}{c | l  l | l l }
             & \multicolumn{2}{c|}{$\pc$}     & \multicolumn{2}{c}{$\nu$} \\
   \hline
   $\alpha$  & $I=I_a$ &  $I=I_b$ &  $I=I_a$  &  $I=I_b$ \\
   \hline
4.0   & 0.20363(1)   & 0.203617(9)  & 1.255(4)     & 1.248(4)     \\
3.8   & 0.21007(1)   & 0.210075(9)  & 1.253(6)     & 1.242(5)     \\
3.6   & 0.218241(9)  & 0.218239(9)  & 1.257(5)     & 1.249(5)     \\
3.5   & 0.22320(1)   & 0.22318(1)   & 1.249(6)     & 1.239(5)     \\
3.4   & 0.228815(7)  & 0.228803(7)  & 1.245(5)     & 1.240(4)     \\
3.2   & 0.24262(1)   & 0.24261(1)   & 1.249(6)     & 1.247(5)     \\
3.1   & 0.25113(2)   & 0.25110(2)   & 1.246(8)     & 1.249(8)     \\
3.0   & 0.26089(1)   & 0.26086(1)   & 1.259(5)     & 1.256(7)     \\
2.9   & 0.27219(2)   & 0.27215(2)   & 1.270(8)     & 1.270(6)     \\
2.85  & 0.27847(1)   & 0.27844(1)   & 1.286(6)     & 1.280(6)     \\
2.8   & 0.28522(1)   & 0.28519(1)   & 1.304(6)     & 1.299(7)     \\
2.75  & 0.29248(2)   & 0.29244(1)   & 1.323(7)     & 1.316(6)     \\
2.7   & 0.30025(1)   & 0.30020(1)   & 1.333(6)     & 1.332(5)     \\
2.6   & 0.317526(8)  & 0.317473(8)  & 1.384(4)     & 1.388(4)     \\
2.5   & 0.33726(1)   & 0.33723(1)   & 1.455(6)     & 1.457(6)     \\
2.4   & 0.35963(2)   & 0.35964(2)   & 1.54(1)      & 1.54(1)      \\
2.3   & 0.38508(7)   & 0.38526(8)   & 1.74(2)      & 1.78(2)      \\
2.2   & 0.4135(2)    & 0.4140(2)    & 2.0(1)       & 2.1(1)       \\
2.1   & 0.4500(7)    & 0.4521(8)    & 3.2(2)       & 3.4(2)       \\
2.0   & 0.502(2)     & 0.514(6)     & 13(5)        & 13(4)       
  \end{tabular}
 \end{ruledtabular}
\end{table}

\subsection{Crossing-point analysis}
\label{subsec:crossing}

\begin{figure}
 \includegraphics[width=0.8420\columnwidth]{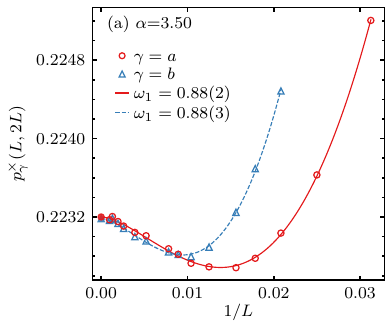}
 \includegraphics[width=0.8420\columnwidth]{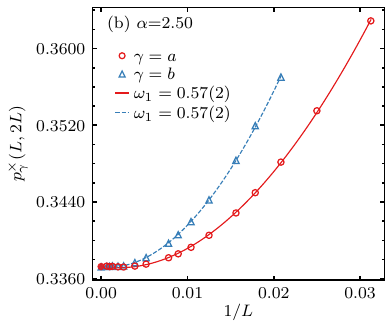}
 \caption{
 \label{fig:px-L}
 System-size dependence of the crossing points 
 $p_{\gamma}^{\times}(L, 2L)$
 for (a) $\alpha=3.5$ and (b) $\alpha=2.5$.
 Open circles (triangles) represent the results for $\gamma=a$ ($b$),
 while solid ones at $1/L=0$ indicate the values of $\pc$ obtained 
 by the data-collapse analysis shown in 
 Figs.~\ref{fig:vsLmin-alpha3500} and 
 \ref{fig:vsLmin-alpha2500}, as well as in Table~\ref{tbl:pc-nu}.
 Solid and dashed lines are independent fits to the data for 
 the two partitioning schemes,  
 $\gamma=a$ and $\gamma=b$, respectively, using 
 the functional from given in Eq.~(\ref{eq:px-L}), 
 where $\pc$ and $\nu$ are fixed 
 at the extrapolated values in the thermodynamic limit 
 obtained by the data-collapse analysis, 
 and $\omega_{2}=2\omega_{1}$ is assumed.
 The fitting values of $\omega_{1}$ are indicated in each panel. 
 }
\end{figure}

\begin{figure}
 \includegraphics[width=0.80\columnwidth]{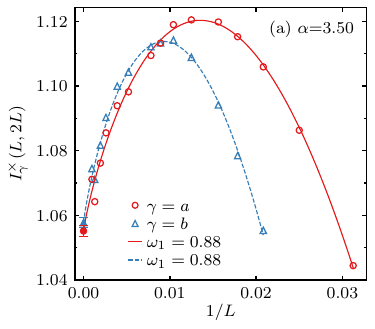}
 \includegraphics[width=0.80\columnwidth]{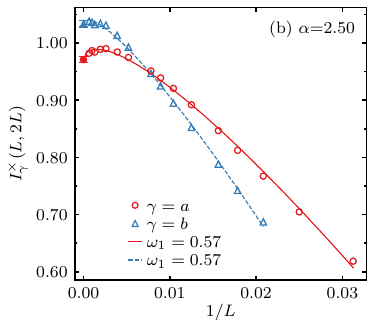}
 \caption{
 \label{fig:Ix-L}
 System-size dependence of the QCMI 
 at the crossing points for (a) $\alpha=3.5$ and (b) $\alpha=2.5$.
 Open circles (triangles) represent the results for 
 $\gamma=a$ ($b$).
 Solid and dashed lines are fits to the data using the functional 
 form given in Eq.~(\ref{eq:Ix-L}),
 where $\omega_{1}$ is fixed to the values obtained 
 from the fits of $p_{\gamma}^{\times}(L, 2L)$ in 
 Fig.~\ref{fig:px-L}, and $\omega_{2}=2\omega_{1}$ is assumed.
 }
\end{figure}

With the accurate estimates of $\pc$ and $\nu$ obtained from 
the data-collapse analysis, we now return to the crossing-point 
analysis. Crossing points are defined as the values of $p$ and 
$I_{\gamma}$ at which the curves of $I_{\gamma}(p, L)$ and 
$I_{\gamma}(p, 2L)$ intersect. These points are denoted as 
$p_{\gamma}^{\times}(L, 2L)$ and $I_{\gamma}^{\times}(L, 2L)$, 
respectively. 
A key difficulty in this analysis is the nonmonotonic finite-size 
behavior of the crossing points, as clearly illustrated in 
Figs.~\ref{fig:px-L}(a) and \ref{fig:Ix-L}(a). 
To capture this behavior, it is necessary to consider at least 
two correction terms in the scaling forms:  
\begin{equation}
 p_{\gamma}^{\times}(L, 2L) = \pc         + L^{-1/\nu} \left( a_{1} L^{-\omega_{1}} + a_{2} L^{-\omega_{2}} \right)
\label{eq:px-L}
\end{equation}
and 
\begin{equation}
 I_{\gamma}^{\times}(L, 2L) = \frac{c}{3} + b_{1} L^{-\omega_{1}} + b_{2} L^{-\omega_{2}},
\label{eq:Ix-L}
\end{equation}
where $a_{i}$ and $b_{i}$ ($i$=1,2) are constants, 
and the correction exponents are ordered such that 
$\omega_{2}>\omega_{1}>0$. 
Note that $I_{b}^{\times}(L, 2L)$ is normalized by a constant factor 
($\simeq 1.83$) 
so that its thermodynamic limit matches that of 
$I_{a}^{\times}(L, 2L)$, i.e., $c/3$, 
[see Eqs.~(\ref{eq:IACBa}) and (\ref{eq:IACBb})]. 
There are two possible scenarios for the origin of the second 
correction term~\cite{Ma-PRL2018}.
One possibility is that it arises from a second irrelevant scaling 
field, independent of the first. 
The other is that it corresponds to the second higher-order 
term of the same irrelevant filed as the first one, 
i.e, $\omega_{2}=2\omega_{1}$. 
Although the scaling forms in Eqs.~(\ref{eq:px-L}) 
and (\ref{eq:Ix-L}) appear relatively simple, 
we find that fitting both correction terms with sufficient 
precision is quite challenging~\cite{Ma-PRL2018,Shao-S2016}. 
This difficulty persists even when we fix $\pc$ and $\nu$ 
in Eq.~(\ref{eq:px-L}) to the values obtained from the 
data-collapse analysis in Table~\ref{tbl:pc-nu}. 

As a result, we are unable to reliably test the first scenario 
involving two independent correction exponents 
by treating $\omega_{1}$ and $\omega_{2}$ as free parameters.
Instead, motivated by the second scenario, we achieve reasonable 
fits by fixing $\pc$ and $\nu$, and imposing 
$\omega_{2}=2\omega_{1}$ in Eq.~(\ref{eq:px-L}), 
as shown in Fig.~\ref{fig:px-L}. 
It is worth noting that the fitted values of $\omega_{1}$ are 
consistent between $\gamma=a$ and $\gamma=b$ 
(see Table~\ref{tbl:w1-c0} for various values of $\alpha$). 
This consistency strongly supports the second scenario.
Moreover, the obtained values of $\omega_{1}$ are compatible with 
the cross-point data of 
$I_{\gamma}^{\times}(L, 2L)$ when fitted to Eq.~(\ref{eq:Ix-L}), 
where only $c$ and $b_{i}$ are fitting parameters, as shown 
in Fig.~\ref{fig:Ix-L}. 
We also find that the obtained values of $c$ are consistent 
between the two different partitioning schemes 
at least for $3.0 \lesssim \alpha$ (see Table~\ref{tbl:w1-c0}), 
as will be discussed in the next section. This further supports 
the assumption $\omega_{2}=2\omega_{1}$.

\begin{table}[tb]
 \caption{\label{tbl:w1-c0}%
 Summary of the correction exponent $\omega_{1}$ and 
 the coefficient $c/3$ obtained from the 
 crossing-point analysis of $p_{\gamma}^{\times}(L, 2L)$ 
 and $I_{\gamma}^{\times}(L, 2L)$, assuming 
 $\omega_{2}=2\omega_{1}$.
 The numbers in parentheses indicate the uncertainty in the last 
 digits.
 }
 \begin{ruledtabular}
  \begin{tabular}{c | l  l | l l }
             & \multicolumn{2}{c|}{$\omega_{1}$}     & \multicolumn{2}{c}{$c/3$} \\
   \hline
   $\alpha$  & $I=I_a$ &  $I=I_b$ &  $I=I_a$  &  $I=I_b$ \\
   \hline
4.0   & 0.94(5)      & 0.92(7)           & 1.053(2)     & 1.052(2)     \\
3.8   & 0.93(4)      & 0.85(5)           & 1.050(3)     & 1.044(3)     \\
3.6   & 0.94(3)      & 0.93(4)           & 1.058(1)     & 1.058(2)     \\
3.5   & 0.88(2)      & 0.88(3)           & 1.055(2)     & 1.058(2)     \\
3.4   & 0.90(1)      & 0.85(3)           & 1.061(1)     & 1.061(2)     \\
3.2   & 0.83(2)      & 0.80(4)           & 1.056(1)     & 1.059(2)     \\
3.1   & 0.76(2)      & 0.76(4)           & 1.049(2)     & 1.057(2)     \\
3.0   & 0.76(2)      & 0.75(3)           & 1.048(1)     & 1.059(3)     \\
2.9   & 0.70(2)      & 0.70(3)           & 1.034(2)     & 1.053(3)     \\
2.85  & 0.73(2)      & 0.68(2)           & 1.035(2)     & 1.050(3)     \\
2.8   & 0.68(2)      & 0.67(3)           & 1.021(3)     & 1.044(4)     \\
2.75  & 0.67(2)      & 0.65(3)           & 1.010(4)     & 1.039(5)     \\
2.7   & 0.65(2)      & 0.67(3)           & 1.004(4)     & 1.044(5)     \\
2.6   & 0.61(2)      & 0.60(3)           & 0.984(4)     & 1.030(5)     \\
2.5   & 0.57(2)      & 0.57(2)           & 0.971(5)     & 1.034(5)     \\
2.4   & 0.55(1)      & 0.53(1)           & 0.987(7)     & 1.060(8)     \\
2.3   & 0.51(1)      & 0.46(1)           & 1.025(8)     & 1.123(9)     \\
2.2   & 0.52(1)      & 0.44(2)           & 1.093(7)     & 1.24(1)      \\
2.1   & 0.43(2)      & 0.41(2)           & 1.228(8)     & 1.394(8)     \\
2.0   & 0.55(3)      & 0.8(2)            & 1.01(1)      & 0.84(2)      \\
  \end{tabular}
 \end{ruledtabular}
\end{table}

\section{Discussion}\label{sec:discussion}

\begin{figure}
 \centering
 \includegraphics[width=0.85\columnwidth]{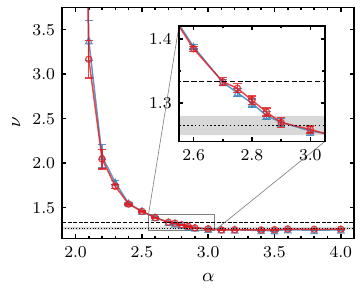}
 \caption{
 \label{fig:nu-alpha}
 $\alpha$-dependence of the correlation-length critical exponent 
 $\nu$.
 The dotted line with the gray shaded area represents the result of 
 the nearest-neighbor model, $\nu=1.265(15)$~\cite{Sierant-PRB2022}.
 The inset shows a magnified view near $\alpha=3.0$.
 For comparison, the prediction of classical two-dimensional 
 percolation theory, $\nu=4/3$, is also indicated by the dashed line.
 }
\end{figure}

In Fig.~\ref{fig:nu-alpha}, we plot the correlation-length exponent 
$\nu$ estimated from the data-collapse 
analysis (see Table~\ref{tbl:pc-nu}) as a function of $\alpha$. 
We confirm that for $\alpha \gtrsim 3$, our model belongs to the 
same universality class as that of the short-range interacting 
models~\cite{Block-PRL2022,Sahu-PRB2022}.
In particular, our numerical estimates are in full agreement with 
recent results for the model with nearest-neighbor 
Clifford
gates~\cite{Sierant-PRB2022}, and differ significantly from 
the value predicted by classical two-dimensional percolation theory, 
$\nu=4/3$.

\begin{figure}
 \centering
 \includegraphics[width=0.85\columnwidth]{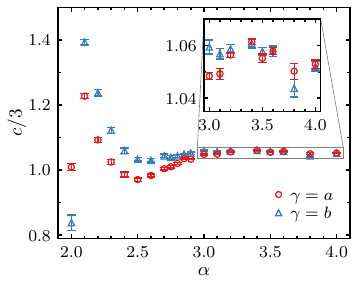}
 \caption{
 \label{fig:c0-alpha}
 Coefficient $c/3$ of the logarithmic term in the entanglement entropy as a function of $\alpha$.
 Circles and triangles represent the results for $\gamma=a$ and 
 $\gamma=b$, respectively.
 The inset shows a magnified view of the region 
 $3 \leqslant \alpha \leqslant 4$.
 }
\end{figure}

The existence of a universality class behavior 
for $\alpha \gtrsim 3$ is 
further supported by the coefficient $c$ of the logarithmic term 
in the entanglement entropy, as shown in Fig.~\ref{fig:c0-alpha}.
The close agreement of $c$ values estimated from the two 
different partitioning schemes, $\gamma=a$ and $b$, 
strongly supports the presence of emergent conformal criticality 
as described by Eqs.~(\ref{eq:Sx})--(\ref{eq:IACBb}) for 
$\alpha \gtrsim 3$, in line with the behavior observed in the 
short-range interacting models.
In contrast, the noticeable discrepancy in the estimated values of  
$c$ between the two partitioning schemes for $\alpha \lesssim 3$ 
indicates a breakdown of the scaling relations in 
Eqs.~(\ref{eq:Sx})--(\ref{eq:IACBb}). 
This deviation from emergent conformal criticality 
corroborates previous findings of non-conformal-field-theory 
criticality, as evidenced by the deviation of the dynamical critical 
exponent $z$ from unity in Ref.~\cite{Block-PRL2022}. 
It is worth noting, however, that the critical point $p_c$ and 
the exponent $\nu$ remain consistent between the two 
partitioning schemes for all values of 
$\alpha$ (see Table~\ref{tbl:pc-nu}). This consistency 
highlights that the QCMI remains a valid and robust 
probe of the measurement-induced entanglement phase 
transition, even in the absence of emergent conformal criticality.

Now we compare the coefficient $c/3$ of the logarithmic term 
in the entanglement entropy with those obtained for monitored 
nearest-neighbor random Clifford circuits 
in Refs.~\cite{Li-PRB2019a,Zabalo-PRB2020,Sierant-PRB2022}. 
While our analysis uses the base-2 logarithm, previous studies 
employed the natural logarithm in defining this coefficient. 
To enable direct comparison, we convert our result as 
\begin{equation}
    \tilde{c} = \frac{1}{\ln 2} \times \frac{c}{3} \simeq 1.443 \times \frac{c}{3}, 
\end{equation}
so that $\tilde{c}$ is consistent with the convention used in the 
earlier works. 
From our estimate of $c/3 = 1.053(2)$ for the shortest-range case 
$\alpha =4 $ using the partitioning scheme $\gamma=a$ 
(see Table~\ref{tbl:w1-c0}), we obtain $\tilde{c} = 1.519(3)$. 
This value is noticeably smaller than those in the previous 
studies: 
$\tilde{c} = 1.6$ with up to 512 qubits~\cite{Li-PRB2019a},
$\tilde{c} = 1.61(3)$ with up to 512 qubits~\cite{Zabalo-PRB2020}, 
and $\tilde{c} = 1.57(1)$ with up to 10240 
qubits~\cite{Sierant-PRB2022} for monitored nearest-neighbor 
random Clifford circuits. 
Furthermore, we find that $c/3$ exhibits only minor variation 
across different partitioning schemes ($\gamma=a,b$) and 
for a range of $\alpha \gtrsim 3$ 
[see the inset of Fig.~\ref{fig:c0-alpha} and Table~\ref{tbl:w1-c0}]. 
This supports the conclusion that our estimate of $\tilde{c}$ 
remains consistently smaller than those in the previous works, 
even when accounting for uncertainties.

\begin{figure*}
 \centering
\includegraphics[width=0.66\columnwidth]{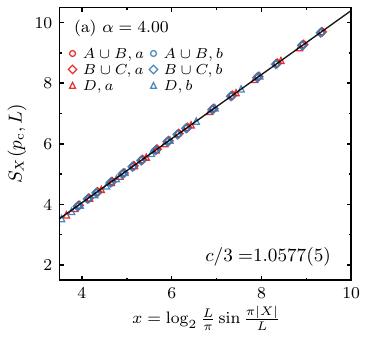} 
\includegraphics[width=0.66\columnwidth]{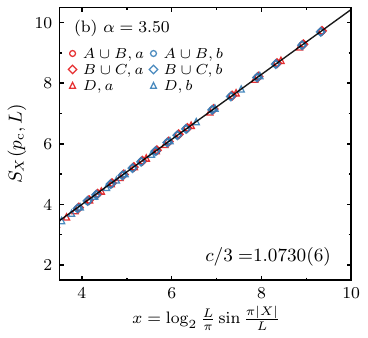}
\includegraphics[width=0.66\columnwidth]{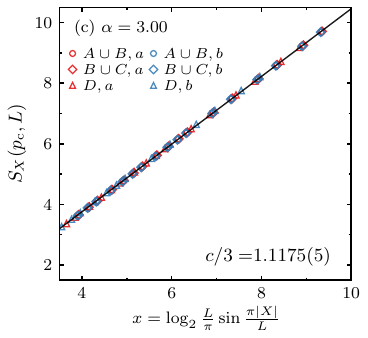} 
 \caption{
 \label{fig:Sx-alpha}
  Entanglement entropy at criticality as a function of chord length 
  for 
  (a) $\alpha=4.0$,
  (b) $\alpha=3.5$, and
  (c) $\alpha=3.0$. 
   Results are shown for various subsystems, $X=A\cup B$, $B \cup C$, and $D=\overline{A\cup B\cup C}$, under partitioning schemes 
   $\gamma=a$ and $b$, using different symbols and colors.
  Solid lines represent fits to the data using the form given 
  in Eq.~(\ref{eq:Sx}). 
  The estimated values of $c/3$ are also indicated in each panel. 
 }
\end{figure*}

To clarify the origin of this discrepancy, we also estimate 
$\tilde{c}$ from the entanglement entropy at criticality, 
as shown in Fig.~\ref{fig:Sx-alpha}. Using the critical point 
$\pc$ obtained from the data-collapse analysis, 
we fit the entanglement entropy of the form given in 
Eq.~(\ref{eq:Sx}), treating $c/3$ and $c'$ as fitting parameters. 
We obtain $\tilde{c}=1.6126(7)$, $1.5484(9)$, and $1.5263(7)$
[$c/3=1.1175(5)$, $1.0730(6)$, and $1.0577(5)$] 
for $\alpha=3.0$, $3.5$, and $4.0$, respectively. 
These results indicate that $c/3$ obtained from the entanglement 
entropy approaches the value estimated from the QCMI from above 
as $\alpha$ increases (see Supplemental Material~\cite{suppl} for 
results at other values of $\alpha$). 
Since finite-size effects are expected to be smaller 
at larger $\alpha$, the estimates of $\tilde{c}$ from the 
entanglement entropy at criticality are more reliable for larger 
$\alpha$ than for the smaller ones. Nevertheless, 
we argue that the estimates of $\tilde{c}$ based on the QCMI 
are generally more accurate, as the QCMI enables crossing-point 
analyses by eliminating the logarithmic dependence at criticality, 
as shown in Eqs.~(\ref{eq:IACBa}) and (\ref{eq:IACBb}).
Indeed, the values of $\tilde{c}$ estimated via the QCMI exhibit 
much weaker dependence on $\alpha$ [see the inset of 
Fig.~\ref{fig:c0-alpha} and Table~\ref{tbl:w1-c0}] compared to 
those obtained from the entanglement entropy~\cite{suppl}. 
We therefore adopt $\tilde{c} = 1.519(3)$, obtained from the QCMI 
for $\alpha=4.0$, as our final estimate for the shortest-range 
interacting case.

\section{Conclusion}~\label{sec:conclusion}

Using stabilizer circuits with variable-range two-qubit Clifford 
gates and local projective measurements, we have demonstrated that 
the QCMI, $I(A:C|B)$, serves as a powerful probe of the 
measurement-induced entanglement phase transition. 
%
An important advantage of the QCMI is its ability to not only 
distinguish between volume-law and area-law entanglement phases, 
but also to provide access to the universal coefficient $c$ 
of the logarithmic term in the entanglement entropy at criticality 
[see Eq.~(\ref{eq:IACB0})]. 
%
Through careful finite-size scaling analyses, we have 
determined the phase diagram and obtained reliable estimates of 
both the correlation-length critical exponent $\nu$ and the 
universal coefficient $c$. 
Notably, for short-range interacting cases ($\alpha \gtrsim 3$), 
the consistency of the estimated values of $c$ across two distinct 
partitioning schemes, each yielding different values of 
$\frac{1}{3}\log_2\frac{l_{A\cup B}l_{B\cup C}} {l_{B}l_{A\cup B\cup C}}$ in Eq.~(\ref{eq:IACB0}), further validates the robustness of 
our approach. 
Finally, the finite-size scaling analysis scheme based on the QCMI 
presented in this work is not limited to stabilizer circuits. 
Its extension to more general, non-stabilizer circuits remains an 
important direction for future research.

\acknowledgements
A portion of the numerical simulations was performed
using the HOKUSAI supercomputer at RIKEN (Project ID: Q22525) 
and the supercomputer Fugaku installed at RIKEN R-CCS
(Project ID: hp220138). 
This work was supported by Grant-in-Aid for Scientific Research (C) 
(No.~JP18K03475, No.~JP21K03395, No.~JP22K03520, and No.~JP24K06894) 
and Grant-in-Aid for Scientific Research (A) (No.~JP21H04446) from 
MEXT, Japan. 
Additional support was provided by the UTokyo 
Quantum Initiative, the RIKEN TRIP initiative (RIKEN Quantum), 
and the COE research grant in 
computational science from Hyogo Prefecture and Kobe City through 
Foundation for Computational Science.
We also acknowledge funding from JST COI-NEXT (Grant No. JPMJPF2221) 
and 
the Program for Promoting Research on the Supercomputer Fugaku 
(Grant No. MXP1020230411) from MEXT, Japan.

\bibliography{MIPT}
\end{document}